\documentclass[pdftex,twocolumn,epjc3]{svjour3}          

\RequirePackage[T1]{fontenc}
\smartqed  
\RequirePackage{graphicx}
\RequirePackage{newtxtext,newtxmath}      
\RequirePackage{flushend}
\RequirePackage[numbers,sort&compress]{natbib}
\RequirePackage[colorlinks,citecolor=blue,urlcolor=blue,linkcolor=blue]{hyperref}
\usepackage{amsmath}
\usepackage{xspace}
\usepackage[dvipsnames]{xcolor}
\usepackage{enumitem}
\usepackage{booktabs}
\usepackage[referable]{threeparttablex}
\usepackage{todonotes}
\usepackage{comment}
\usepackage[tight]{subfigure}
\usepackage[symbol]{footmisc}
\usepackage{units}
\renewlist{tablenotes}{enumerate}{1}
\makeatletter
\setlist[tablenotes]{label=\tnote{\alph*},ref=\alph*,itemsep=\z@,topsep=\z@skip,partopsep=\z@skip,parsep=\z@,itemindent=\z@,labelindent=\tabcolsep,labelsep=.2em,leftmargin=*,align=left,before={\footnotesize}}
\makeatother

\newcommand{\nue}{\ensuremath{\nu_{e}}\xspace}
\newcommand{\nuebar}{\ensuremath{\overline{\nu}_{e}}\xspace}
\newcommand{\numubar}{\ensuremath{\overline{\nu}_{\mu}}\xspace}
\newcommand{\numu}{\ensuremath{\nu_{\mu}}\xspace}

\newcommand{\newtext}[1]{\textcolor{black}{#1}\xspace}

\journalname{Eur. Phys. J. C}

\begin{document}
\title{Direct comparison of sterile neutrino constraints from cosmological data, \nue disappearance data and $\numu\rightarrow\nue$ appearance data in a $3+1$ model}

\author{
Matthew Adams\thanksref{Manchester} \and Fedor Bezrukov\thanksref{Manchester} \and Jack Elvin-Poole\thanksref{Ohio} \and Justin J. Evans\thanksref{Manchester} \and Pawel Guzowski\thanksref{Manchester} \and Br\'{i}an \'{O} Fearraigh\thanksref{Nikhev,Amsterdam} \and Stefan S\"oldner-Rembold\thanksref{Manchester}
}
\institute{The University of Manchester, Department of Physics and Astronomy, Manchester, M13 9PL, United Kingdom \label{Manchester} 
\and
Centre for Cosmology and Astro-Particle Physics, and Department of Physics, The Ohio State University, Columbus, OH 43210, USA \label{Ohio}
\and
Nikhef, National Institute for Subatomic Physics, PO Box 41882, Amsterdam, 1009 DB, The Netherlands \label{Nikhev}
\and
University of Amsterdam, Institute of Physics/IHEF, PO Box 94216, Amsterdam, 1090 GE, The Netherlands \label{Amsterdam}}

\date{Received: February 22, 2020 / Accepted: date}
\maketitle

\begin{abstract}
We present a quantitative, direct comparison of constraints on sterile neutrinos derived from neutrino oscillation experiments and from Planck data, interpreted assuming standard cosmological evolution. We extend a $1+1$ model, which is used to compare exclusions contours at the 95\% CL derived from Planck data to those from $\nu_{e}$-disappearance measurements, to a $3+1$ model. This allows us to compare the Planck constraints with those obtained through $\nu_{\mu}\rightarrow\nu_{e}$ appearance searches, which are sensitive to more than one active-sterile mixing angle. We find that the cosmological data fully exclude the allowed regions published by the LSND, MiniBooNE and Neutrino-4 collaborations, and those from the gallium and rector anomalies, at the 95\% CL. Compared to the exclusion region from the Daya Bay $\nu_{e}$-disappearance search, the Planck data are more strongly excluding above $|\Delta m^{2}_{41}|\approx 0.1\,\mathrm{eV}^{2}$ and $m_\mathrm{eff}^\mathrm{sterile}\approx 0.2\,\mathrm{eV}$, with the Daya Bay exclusion being stronger below these values. Compared to the combined Daya Bay/Bugey/MINOS exclusion region on $\nu_{\mu}\rightarrow\nu_{e}$ appearance, the Planck data is more strongly excluding above $\Delta m^{2}_{41}\approx  5\times 10^{-2}\,\mathrm{eV}^{2}$, with the exclusion strengths of the Planck data and the Daya Bay/Bugey/MINOS combination becoming comparable below this value.
\end{abstract}
\section{Introduction}
\label{sec:introduction}

The LSND~\cite{ref:LSND}, MiniBooNE~\cite{ref:MiniBooNENew}, 
and Neutrino-4~\cite{ref:neutrino4} collaborations have made
observations consistent with anomalous neutrino flavour oscillations. Other, related anomalies have been measured
with gallium detectors~\cite{ref:Gallium} and reactor neutrinos~\cite{ref:ReactorAnomaly}.
These observations 
suggest that additional neutrino flavours may exist at a mass scale of $\mathcal{O}(\unit[1]{eV})$, beyond the three flavours of the Standard Model. 

Measurements of the decay width of the $Z$ boson~\cite{ref:ZWidth} conclusively show that only three neutrino flavours with $m_{\nu}<m_{Z}/2$ couple through the weak interaction; these three flavours are termed ``active'', and any additional flavours are therefore referred to as ``sterile''.
The existence of a sterile neutrino can have observable effects since neutrino oscillations allow the sterile flavour states to mix with the active flavour states. Such mixing occurs as the neutrino mass eigenstates are related to the flavour eigenstates through a mixing matrix, the PMNS matrix~\cite{ref:Pontecorvo1,ref:Pontecorvo2,ref:MNS}. The minimal phenomenological $3+1$ model of sterile neutrinos adds a single sterile flavour state and a fourth mass eigenstate.

Limits on the existence of sterile neutrinos have been set by observations of the cosmic microwave background (CMB)~\cite{ref:PlanckSterile} and by numerous neutrino oscillation experiments~\cite{ref:MINOS1,ref:MINOS2,ref:DayaBay,ref:MINOSDayaBay,ref:IceCubeHE,ref:IceCubeLE,ref:NEOS}. In a commonly used model, cosmological measurements set limits on the parameter $\Delta N_{\mathrm{eff}}$, the additional number of relativistic degrees of freedom in the universe arising from the additional neutrino states, and $m_{\mathrm{eff}}^{\mathrm{sterile}}$, the effective mass of the sterile neutrino. In a $3+1$ model, neutrino oscillation experiments set limits on the mass splitting $\Delta m^{2}_{41}=m_{4}^{2}-m_{1}^{2}$, the difference between the squared masses of the additional, fourth mass eigenstate and the lightest neutrino eigenstate, along with the elements of the $4\times 4$ PMNS matrix.

\newtext{Several previous studies~\cite{ref:Hannestad2012,ref:Gariazzo2012,ref:Bergstrom2014,ref:Hannestad2015} have made quantitative comparisons of cosmological and neutrino-oscillation limits on sterile neutrinos. For reviews of the field see, for example, Refs.~\cite{ref:GiuntiReview,ref:BoserReview,ref:KangReview}.  Such comparisons} are complicated due to this difference in parameterization. In a previous article~\cite{ref:Us}, a comparison using a phenomenological model in which only the muon-neutrino flavour mixes into the fourth mass eigenstate was
presented. This $1+1$ model allows only comparisons of \numu disappearance measurements to the cosmological limits. Other studies~\cite{ref:CanadianPaper} have investigated the situation in which only the electron-neutrino flavour is assumed to mix into the fourth mass eigenstate. \newtext{Studies~\cite{ref:mma,ref:Mirizzi,ref:SavianoMirizzi,ref:Gariazzo} are now extending the treatment to the full $3+1$ model that is favoured for phenomenological interpretations of sterile neutrino searches.
In this article, we extend beyond our previous work in~\cite{ref:Us} to the $3+1$ model, to allow a direct comparison of cosmological limits to the LSND and MiniBooNE $\numubar\rightarrow\nuebar$ and $\numu\rightarrow\nue$ observations, showing the comological limits in the parameter space used by LSND and MiniBooNE, and also showing the LSND and MiniBooNE allowed regions in the parameter space of cosmological limits. In doing this, we develop a novel method that allows us to extend our comparisons into the degenerate region in which the sterile mass-splitting $\Delta m^{2}_{41}$ becomes equal to the mass splitting $\Delta m^{2}_{31}$.}


\section{Sterile neutrinos in oscillation experiments}
\label{sec:neutrinooscillations}

In the $3+1$ model, four neutrino flavour eigenstates, $\nu_{l}$ ($l=e,\mu,\tau,s)$, are related to four neutrino mass eigenstates, $\nu_{i}$ ($i=1,2,3,4$), with masses $m_{i}$, by a $4\times4$ extension of the PMNS matrix, $U$:
\begin{equation}
\left|\nu_{l}\right>=\sum_{i=1}^{4}U_{li}\left|\nu_{i}\right>.
\end{equation}
Throughout this paper, we assume all neutrino and antineutrino oscillation probabilities are equal and therefore use the symbol $\nu$ to also refer to $\overline{\nu}$. If a neutrino of energy $E$ is produced in a flavour eigenstate $\nu_{l}$, the probability that it is detected in flavour eigenstate $\nu_{l^{\prime}}$ after traveling a distance $L$ is
\begin{equation}
P_{\nu_{l}\rightarrow\nu_{l^{\prime}}} = \left|
\sum_{i=1}^{4}U_{li}U^{*}_{l^{\prime}i}e^{-im_{i}^{2}L/2E}
\right|^{2}.
\end{equation}
An experiment searching for \nue or \numu disappearance thus measures
\begin{equation}
1-P_{\nu_{l}\rightarrow\nu_{l}} = 4\sum_{i=1}^{3}\sum_{j>i}^{4}\left|U_{li}\right|^{2}\left|U_{lj}\right|^{2}\sin^{2}\left(\frac{\Delta m^{2}_{ji}L}{4E}\right),
\label{eq:dis}
\end{equation}
where $\Delta m^{2}_{ji}=m^{2}_{j}-m^{2}_{i}$ are the mass splittings.
Each mass splitting therefore defines an observable oscillation wavelength, with the elements of the PMNS matrix governing the amplitudes of those oscillations.

Over the majority of the parameter space relevant to sterile-neutrino searches, $|\Delta m^{2}_{41}|\gg|\Delta m^{2}_{31}|>|\Delta m^{2}_{21}|$. Thus, we can choose $L$ and $E$ to probe only the oscillations at the $\Delta m^{2}_{41}$ wavelength, allowing us to approximate the disappearance probabilities 
in Eq.~\ref{eq:dis} to
\begin{eqnarray}
    1-P_{\nu_{e}\rightarrow\nu_{e}}&\approx&\sin^{2}(2\theta_{14})\sin^{2}\left(\frac{\Delta m^{2}_{41}L}{4E}\right),\\
    1-P_{\nu_{\mu}\rightarrow\nu_{\mu}}&\approx&\sin^{2}(2\theta_{24})\sin^{2}\left(\frac{\Delta m^{2}_{41}L}{4E}\right).
\end{eqnarray}
Here, we have introduced the mixing angles $\theta_{ij}$ that are used to parameterize the PMNS matrix. We refer to this approximation of the oscillation probabilities as a $1+1$ model since it assumes only one mass splitting, neglecting the effects of $\Delta m^{2}_{31}$ and $\Delta m^{2}_{21}$, and assuming only one flavour state at a time (either electron or muon) mixes into the fourth mass eigenstate. The mixing angle $\theta_{14}$ quantifies how much electron flavour mixes into the fourth mass eigenstate, and the angle $\theta_{24}$ quantifies this mixing for the muon flavour. In this paper, we use the $1+1$ model for an analysis of $\nue$ disappearance.

In our analysis of $\numu\rightarrow\nue$ appearance we use a $3+1$ model, in which there are three independent mass splittings ($\Delta m^{2}_{21}$, $\Delta m^{2}_{31}$ and $\Delta m^{2}_{41}$), six mixing angles ($\theta_{12}$, $\theta_{13}$, $\theta_{23}$, $\theta_{14}$, $\theta_{24}$, and $\theta_{34}$), and three complex phases ($\delta_{13}$, $\delta_{14}$ and $\delta_{34}$). Still, only the angles $\theta_{14}$ and $\theta_{24}$ and the mass-splitting $\Delta m^{2}_{41}$ are relevant to this work. We set $\theta_{34}=\delta_{14}=\delta_{34}=0$, as these parameters have no impact on our results. The remaining oscillation parameters we set to the best-fit values from a global fit~\cite{ref:GlobalFit}, assuming normal mass ordering: $\Delta m^{2}_{21}=\unit[7.50\times 10^{-5}]{eV^{2}}$, $\Delta m^{2}_{21}=\unit[2.524\times 10^{-3}]{eV^{2}}$, $\sin^{2}\theta_{12}=0.306$, $\sin^{2}\theta_{13}=0.02166$, $\sin^{2}\theta_{23}=0.441$, and $\delta_{13}=0$.

We use the exact oscillation formula for our analysis of $\nue$ appearance. Since in the region of large  $\Delta m^{2}_{41}$ the relevant oscillation probability for $\numu\rightarrow\nue$ is, to a good approximation,
\begin{equation}
    P_{\numu\rightarrow\nue} \approx \sin^{2}(2\theta_{14})\sin^{2}\theta_{24}\sin^{2}\left(\frac{\Delta m^{2}_{41}L}{4E}\right),
\label{eq:vacosc}
\end{equation}
we express limits as a function of $\Delta m^{2}_{41}$ and $\sin^{2}(2\theta_{14})\sin^{2}\theta_{24}\equiv\sin^{2}(2\theta_{\mu e})$.

\section{Data from oscillation experiments}
We use data from collaborations that report allowed regions consistent with sterile neutrino oscillations. Such regions have been reported by the LSND, MiniBooNE, and Neutrino-4 collaborations, in addition to the regions allowed by the reactor and gallium anomalies.
We then compare to the exclusion region from Daya Bay, combined with Bugey-3 and MINOS data, which provides stronger exclusion at lower values of the mass of the fourth mass eigenstate, where the sensitivity of the Planck results decreases.

\subsection{LSND}

The Liquid Scintillator Neutrino Detector (LSND) took data from 1993--1998 at the Los Alamos Meson Physics Facility. A \unit[167]{t} liquid scintillator detector was placed \unit[30]{m} away from a stopped-pion source that produced \numubar with energies up to \unit[52.8]{MeV}~\cite{ref:LSNDNIM}. Appearance of \nuebar was observed in the detector with a total excess of $87.9\pm22.4\mathrm{(stat.)}\pm6.0\mathrm{(syst.)}$ \nuebar events above the expected background~\cite{ref:LSND}. To explain this excess through oscillations, a mass splitting $\Delta m^{2}_{41}\gtrsim\unit[0.03]{eV^{2}}$ is required.

We determine the $90\%$ Confidence Level (CL) allowed region by  requiring $\chi^2-\chi^2_{\rm min}=4.605$ between the observed positron energy spectrum and an estimated spectrum. 
The appearance spectrum is simulated with pseudo-experiments, producing a reconstructed neutrino energy from a reconstructed positron energy and angle, and integrating the reconstructed neutrino energy over the same binning as in~Ref.~\cite{ref:LSND}. The true positron energy, $E_{e^+}$, is the difference between the true neutrino energy and the threshold energy of $1.806$~MeV. The $\nuebar\to e^+$
cross section is estimated to be linear in $E_{e^+}$. The reconstructed positron energy is smeared by a Gaussian function of the form $7\%/\sqrt{E_{e^{+}}/52.8\textrm{~MeV}}$, and its angle is Gaussian-smeared by $12^{\circ}$. The distance $L$ that the neutrino has travelled is uniformly spread in the range $[25.85,34.15]$~m, and a $14$~cm Gaussian smearing is applied to produce a reconstructed distance. The flux is determined for pions decaying at rest to $\numubar$, with an $L^{-2}$ weighting applied. The true neutrino energy and distance is used to calculate the oscillated $\nuebar$ flux with Eq.~\ref{eq:vacosc}.

\subsection{MiniBooNE}

The MiniBooNE experiment was an \unit[818]{t} mineral oil Che\-ren\-kov detector~\cite{ref:MiniBooNENIM} \unit[541]{m} away from the neutrino-produc\-tion target of the Booster Neutrino Beam~\cite{ref:BoosterNeutrinoBeam}. The beam could be configured to produce either \numu or \numubar with mean energy of $\approx\!\!\unit[800]{MeV}$. By searching for the appearance of either \nue or \nuebar, the experiment was sensitive to oscillations driven by a similar range of $\Delta m^{2}_{41}$ as LSND. An excess of activity consistent with \nue and \nuebar was observed. 
We use the CL contours from the Collaboration's public data release~\cite{ref:MINIBOONEDataRelease}.

\subsection{Neutrino-4}

The Neutrino-4 experiment~\cite{ref:neutrino4} searches for the disappearance of \nuebar from the SM3 reactor in Russia. A gadolinium-doped liquid scintillator detector is divided into 50 sections that can be placed at various distances, from $6$ to $\unit[12]{m}$, from the reactor core. The data analysis yields an oscillatory pattern to the $\nuebar$ detection rate as a function of $L/E$ that is interpreted in terms of a sterile neutrino with best-fit oscillation parameters $\Delta m^{2}_{41} = \unit[7.34]{eV^{2}}$, $\sin^{2}(2\theta_{14}) = 0.44$. We take the 95\% CL allowed region directly from Ref.~\cite{ref:neutrino4}.

\subsection{Reactor anomaly}

The reactor anomaly, first described in Ref.~\cite{ref:ReactorAnomaly}, is the observation that, with more modern flux calculations, many short-baseline reactor-\nuebar searches show a deficit compared to the expected flux. This observation can be interpreted as \nuebar disappearance due to oscillations involving a sterile neutrino. We use the 95\% CL allowed region calculated in Ref.~\cite{ref:KoppReactorGallium}.

\subsection{Gallium anomaly}

The gallium anomaly, first described in Ref.~\cite{ref:Gallium}, measured the \nue rate from radioactive calibration sources in the SAGE and GALLEX solar-neutrino detectors. A deficit in the measured rate compared to the expectation can be interpreted as \nue disappearance due to oscillations involving a sterile neutrino. We use the 95\% CL allowed region calculated in Ref.~\cite{ref:KoppReactorGallium}.

\subsection{Daya Bay}

The Daya Bay experiment consists of eight gadolinium-doped liquid scintillator detectors that measure the disappearance of electron antineutrinos from the Daya Bay and Ling Ao nuclear power plants in China~\cite{ref:DayaBayNIM}. The arrangement of eight detectors and six reactor cores provides a range of baselines between \unit[358]{m} and \unit[1925]{m}. The Daya Bay experiment was designed to be sensitive to oscillations driven by $\Delta m^{2}_{31}$ and $\theta_{13}$~\cite{ref:DayaBayTheta13}; however, by looking for non-standard \nuebar disappearance, Daya Bay can also search for oscillations driven by $\Delta m^{2}_{41}$ and $\theta_{14}$ in the range $10^{-4} \lesssim \left|\Delta m^{2}_{41}\right| \lesssim \unit[0.1]{eV^{2}}$~\cite{ref:DayaBay}. We use the Daya Bay data release~\cite{ref:DayaBayDataRelease} to recreate the $\chi^{2}$ surface, and follow the prescribed approach~\cite{ref:DayaBayCLs}, based on the CL$_{\mathrm{s}}$ method~\cite{ref:CLsJunk,ref:CLsRead}, to produce the $95\%$ CL exclusion contour.



\subsection{Bugey-3}

The Bugey-3 experiment took data in the early 1990s. The experiment used two lithium-doped liquid-scintillator detectors~\cite{ref:Bugey3NIM} to search for the disappearance of \nuebar at distances of \unit[15]{m}, \unit[40]{m} and \unit[95]{m} from the Bugey nuclear power plant in France~\cite{ref:Bugey3}. The shorter baseline provides sensitivity to sterile neutrinos at a higher range of $\left|\Delta m^{2}_{41}\right|$ compared to Daya Bay.

\subsection{MINOS}
The MINOS experiment used two steel-scintillator calorimeters~\cite{ref:MINOSNIM} to search for the disappearance of muon neutrinos and antineutrinos from the NuMI  beam at Fermilab~\cite{ref:NuMI} at baselines of \unit[1.04]{km} and \unit[735]{km}. MINOS was designed to be sensitive to oscillations driven by $\Delta m^{2}_{31}$ and $\theta_{23}$~\cite{ref:MINOSStandard}. By searching for non-standard \numu and \numubar disappearance at higher energies, it is also sensitive to oscillations driven by the sterile-neutrino parameters $\Delta m^{2}_{41}$ and $\theta_{24}$~\cite{ref:MINOS1}.

\subsection{Combination of Daya Bay, Bugey-3, and MINOS Data}
The Daya Bay limit was combined with that of Bugey-3 and MINOS to produce limits on the parameters $\Delta m^{2}_{41}$ and $\sin^{2}(2\theta_{\mu e})$ that govern $\numu\rightarrow\nue$ appearance~\cite{ref:MINOSDayaBay}. In performing this combination, the analysis of the Bugey-3 data was updated to use a more recent calculation of the neutron lifetime in the cross-section of the inverse-$\beta$ decay process that is used for \nuebar detection. In addition, the ILL+Vogel flux model~\cite{ref:ILL,ref:Vogel} was replaced with the Huber-Mueller model~\cite{ref:Huber,ref:Mueller}.
We use the combined CL$_{\mathrm{s}}$ surface of Ref.~\cite{ref:MINOSDayaBayDataRelease} to reproduce the $95\%$ CL exclusion contour.

\begin{figure*}[htbp]
\begin{center}
\subfigure[]{\includegraphics[width=1.02\columnwidth]{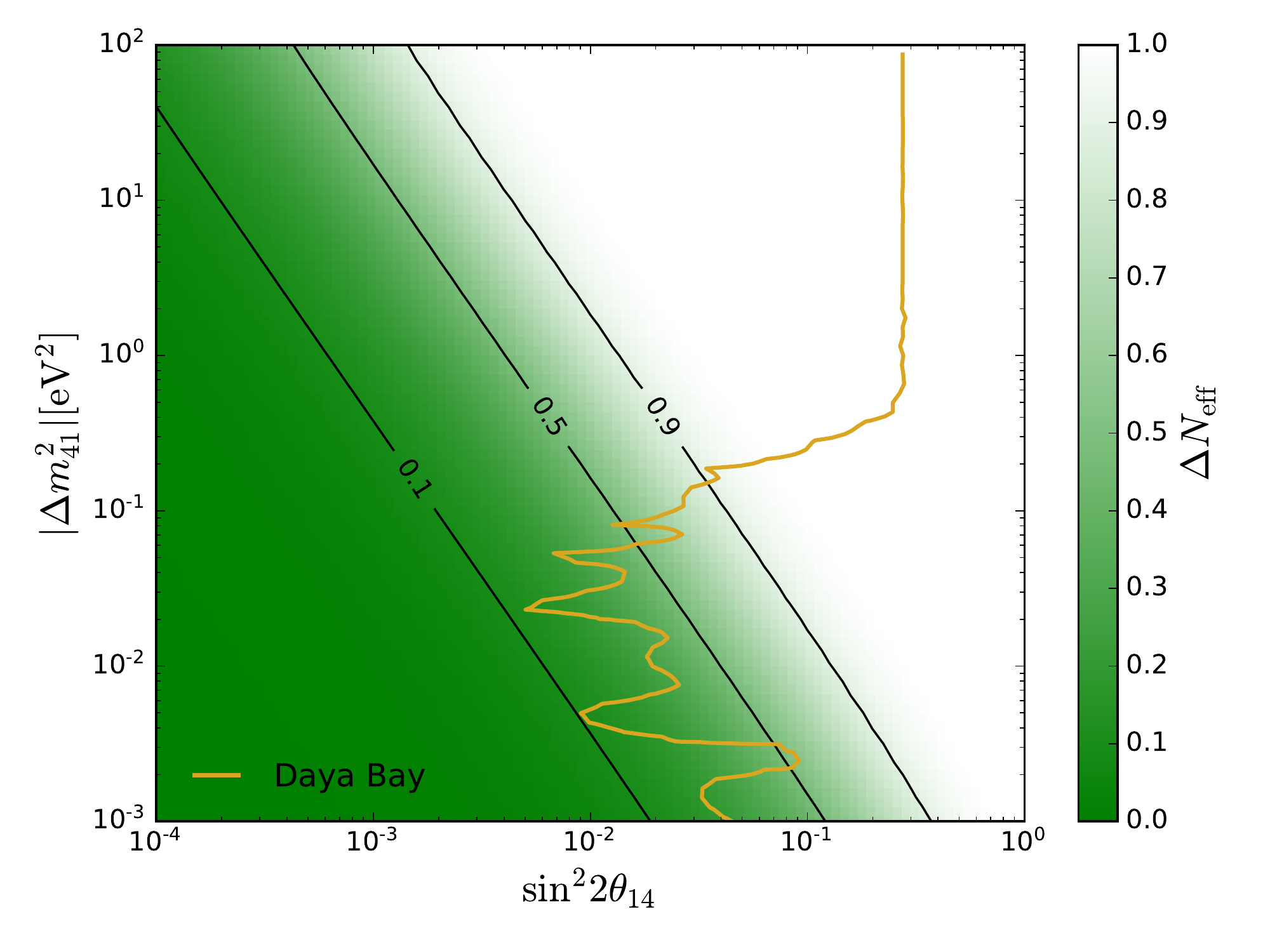}\label{fig:translate_elec_dis_a}}
\subfigure[]{\includegraphics[width=1.02\columnwidth]{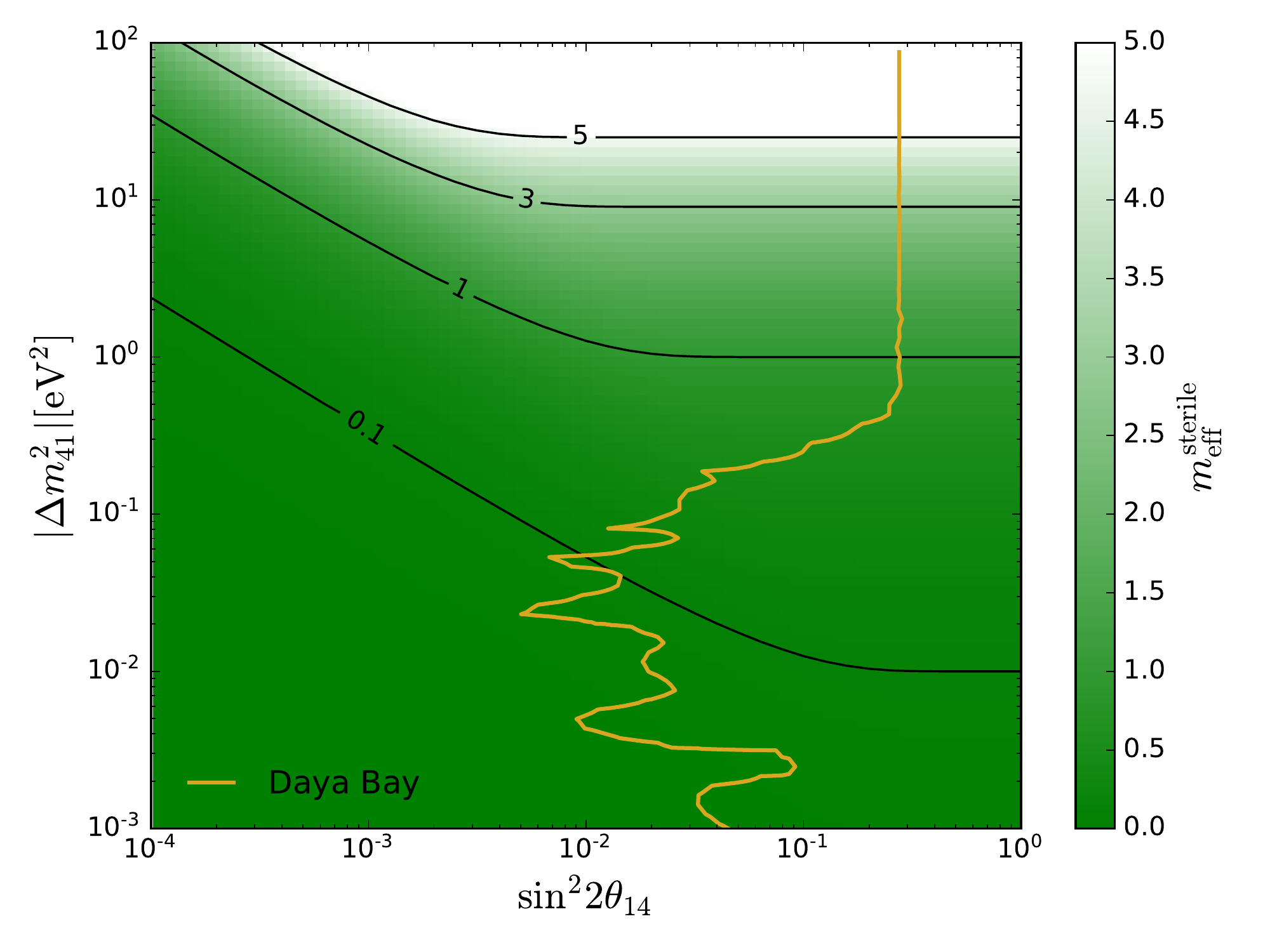}\label{fig:translate_elec_dis_b}}
\subfigure[]{\includegraphics[width=1.02\columnwidth]{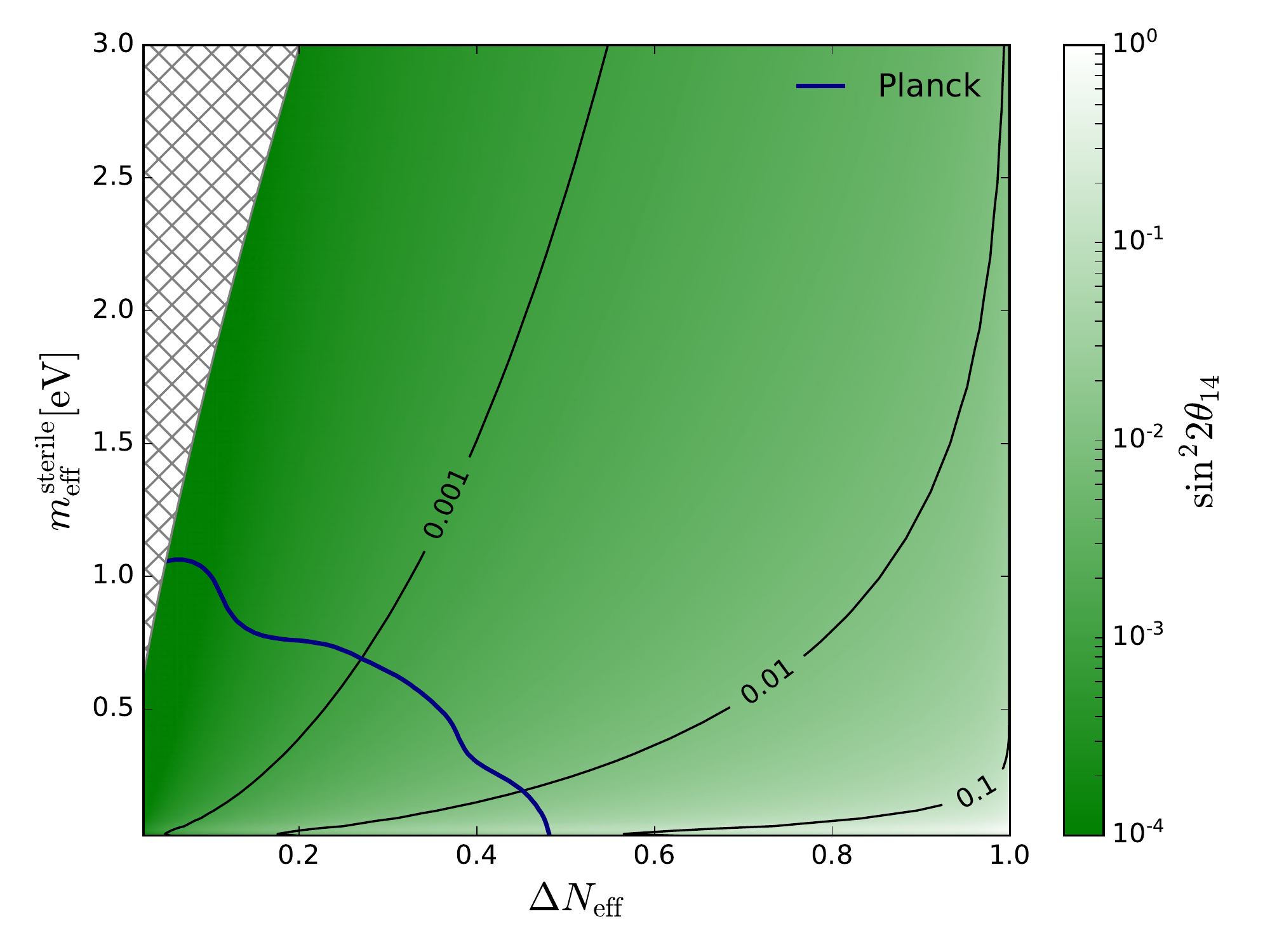}\label{fig:translate_elec_dis_c}}
\subfigure[]{\includegraphics[width=1.02\columnwidth]{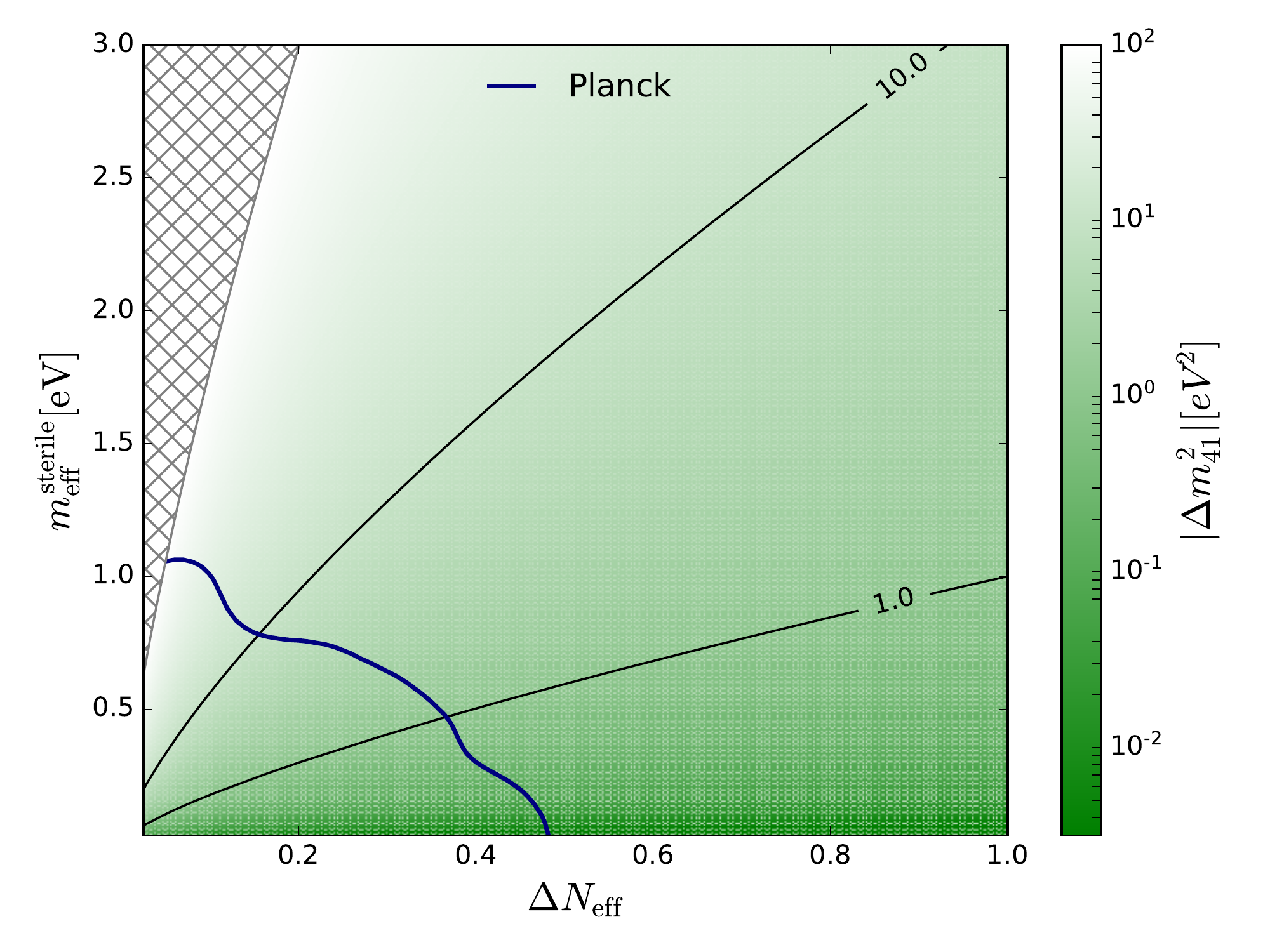}\label{fig:translate_elec_dis_d}}
  \caption{
  (a, b) Cosmological parameters $\Delta N_{\rm eff}^\mathrm{sterile}$ and $m_{\mathrm{eff}}^\mathrm{sterile}$ calculated, using LASAGNA, in the oscillation space of the $1+1$ model that is relevant for \nue and \nuebar disappearance measurements.
  We use the thermal sterile neutrino mass (Eq.~\protect\ref{eq:EffectiveMass}) and assume vanishing lepton asymmetry ($L=0$).  
  We also show the Daya Bay exclusion contour; the region to the right of the contour is ruled out at the $95\%$ CL.
  (c, d) The oscillation parameters of the $1+1$ electron-neutrino disappearance model, $\Delta m^{2}_{41}$ and $\sin^{2} (2 \theta_{14})$, in the cosmological space $(m_{\mathrm{eff}}^{\mathrm{sterile}}$, $\Delta N_{\mathrm{eff}})$. The region above the blue line is excluded by the Planck TT, TE, EE and low-multipole EE power spectra at the $95\%$ CL. A prior of $m_{4} < \unit[10]{eV}$ is applied, shown by the hatched region that has not been considered in our probability density estimation.}
  \label{fig:translations_disappearance}
  \end{center}
\end{figure*}

\section{Sterile neutrinos in cosmological measurements}
\label{sec:cosmology}

The presence of one or more sterile neutrinos can affect the power spectrum of the CMB. The effective mass of the sterile neutrino is defined as
\mbox{$
    m_{\mathrm{eff}}^{\mathrm{sterile}} = 
    \left(94.1\,\Omega_{\mathrm{sterile}}h^{2}\right)\,\mathrm{eV},
$}
where $h=H/100$ with the Hubble parameter $H$, and $\Omega_{\mathrm{sterile}}$ is the contribution of sterile neutrinos to the matter energy-density in the Universe. 
The neutrino number density, $n_{\nu}$, is expressed as a function of the number of effective neutrino species, $N_{\mathrm{eff}}$, as
\begin{equation}
    n_{\nu} = 
    \left(\frac{3}{4}\right) N_{\mathrm{eff}}
    \left(\frac{4}{11}\right)n_{\gamma},
\end{equation}
where $n_{\gamma}$ is the number density of photons in the CMB. 
Standard cosmology predicts $N_{\mathrm{eff}}=3.046$, 
since the process of neutrino decoupling from the CMB was not instantaneous, and neutrinos still interacted with leptons in the primordial plasma~\cite{Abazajian:2013oma}. This allows us to define the effective number of additional radiative degrees of freedom, equivalent to the effective number of additional neutrino species, as \mbox{$
    \Delta N_{\mathrm{eff}} = N_{\mathrm{eff}}-3.046.
$}

We relate $m_{\mathrm{eff}}^{\mathrm{sterile}}$ and the mass of the fourth neutrino mass eigenstate, $m_{4}$ using the standard relationship~\cite{ref:PlanckSterile} 
\begin{equation}
    m_{\mathrm{eff}}^{\mathrm{sterile}} =
    \left(\frac{T_{s}}{T_{\nu}}\right)^{3}m_{4}
    =(\Delta N_{\mathrm{eff}})^{3/4}m_{4}.
    \label{eq:EffectiveMass}
\end{equation}
Here, we assume a thermally distributed sterile neutrino with a temperature $T_{s}$ that may differ from the active neutrino thermalisation temperature $T_{\nu}$

A fully thermalized sterile neutrino with temperature $T_{s}=T_{\nu}$ corresponds to a measured $\Delta N_{\mathrm{eff}}=1$ and $m_{\mathrm{eff}}^{\mathrm{sterile}}=m_{4}$. An alternative relationship between $m_{\mathrm{eff}}^{\mathrm{sterile}}$ and $m_{4}$, the Dodelson-Widrow mechanism~\cite{Dodelson:1993je}, assumes that $\Delta N_{\mathrm{eff}}$ acts as a linear scaling factor, $
    m_{\mathrm{eff}}^{\mathrm{sterile}}=\Delta N_{\mathrm{eff}}m_{4}.$
The choice of this
function does not significantly impact our results.

\section{The Planck experiment}     

The Planck satellite made detailed observations of an\-isot\-ropies of the CMB between 2009 and 2013, over a frequency range from \unit[30] to \unit[857]{GHz}~\cite{ref:PlanckLFI,ref:PlanckHFI}. The Planck Collaboration combines data from the TT, TE and EE power spectra, the low-multipole EE power spectrum (LowE), CMB lensing, and baryon acoustic oscillations (BAO) to set limits of $N_{\mathrm{eff}} < 3.29$ and $m_{\mathrm{eff}}^{\mathrm{sterile}} < \unit[0.23]{eV}$~\cite{ref:PlanckSterile}. 
These results arise from the use of a flat prior in the range $0<m_{\mathrm{eff}}^{\mathrm{sterile}}<10~$eV.
A more restrictive prior results in more constraining limits. A flat prior in the range $0<\Delta N_{\mathrm{eff}}<1$ is also used. The Planck analysis assumes a normal neutrino-mass ordering and active states with masses $m_{1}=m_{2}=0$ and $m_{3}=\unit[0.06]{eV}$.

\newtext{
To obtain these limits on sterile neutrinos, we used data sets provided by the Planck Collaboration. They fit the data using the \texttt{CosmoMC} software~\cite{ref:cosmomc2002,ref:cosmomc2013}, based on a $\Lambda\textrm{CDM}+m_{\mathrm{eff}}^{\mathrm{sterile}}+\Delta N_{\mathrm{eff}}$ model. Neutrino and nuisance parameters are varied to build a large number of points in the parameter space. The cosmological priors used are described in Section 2.1 of Ref~\cite{ref:PlanckSterile}. The Planck Collaboration provides the Markov Chain Monte Carlo (MCMC) points in Ref.~\cite{ref:PlanckDataRelease}.
We derive exclusion limits in the $(\Delta N_{\mathrm{eff}}, m_{\mathrm{eff}}^{\mathrm{sterile}})$ space by using kernel density estimation (implemented in \texttt{scipy}~\cite{ref:scipy}) over the MCMC points to find the most probable point in the two-dimensional space, as well as the region around it that contains $95\%$ of the integrated probability when ordered by probability density.
}
\section{Electron neutrino disappearance in a $1+1$ model}
\label{sec:nuedisappearance}
To
translate from the parameter space ($\Delta N_{\text{eff}}$, $m_{\text{eff}}^{\text{sterile}}$) to the parameter space ($\sin^{2} 2\theta_{14}$, $|\Delta m^{2}_{41}|$), 
we use LASAGNA \cite{ref:Hannestad2013} for
calculating $\Delta N_{\text{eff}}$ as a function of the mass splitting $|\Delta m^{2}_{41}|$ and mixing angle $\sin^{2} 2\theta_{14}$.
LASAGNA solves the quantum kinetic equations describing neutrino thermalization in the early universe by evolving the equations over a temperature range for input values of $|\Delta m^{2}_{41}|$ and $\sin^{2} (2\theta_{14})$.

\begin{figure*}[htbp]
\subfigure[]{\includegraphics[width=0.99\columnwidth]{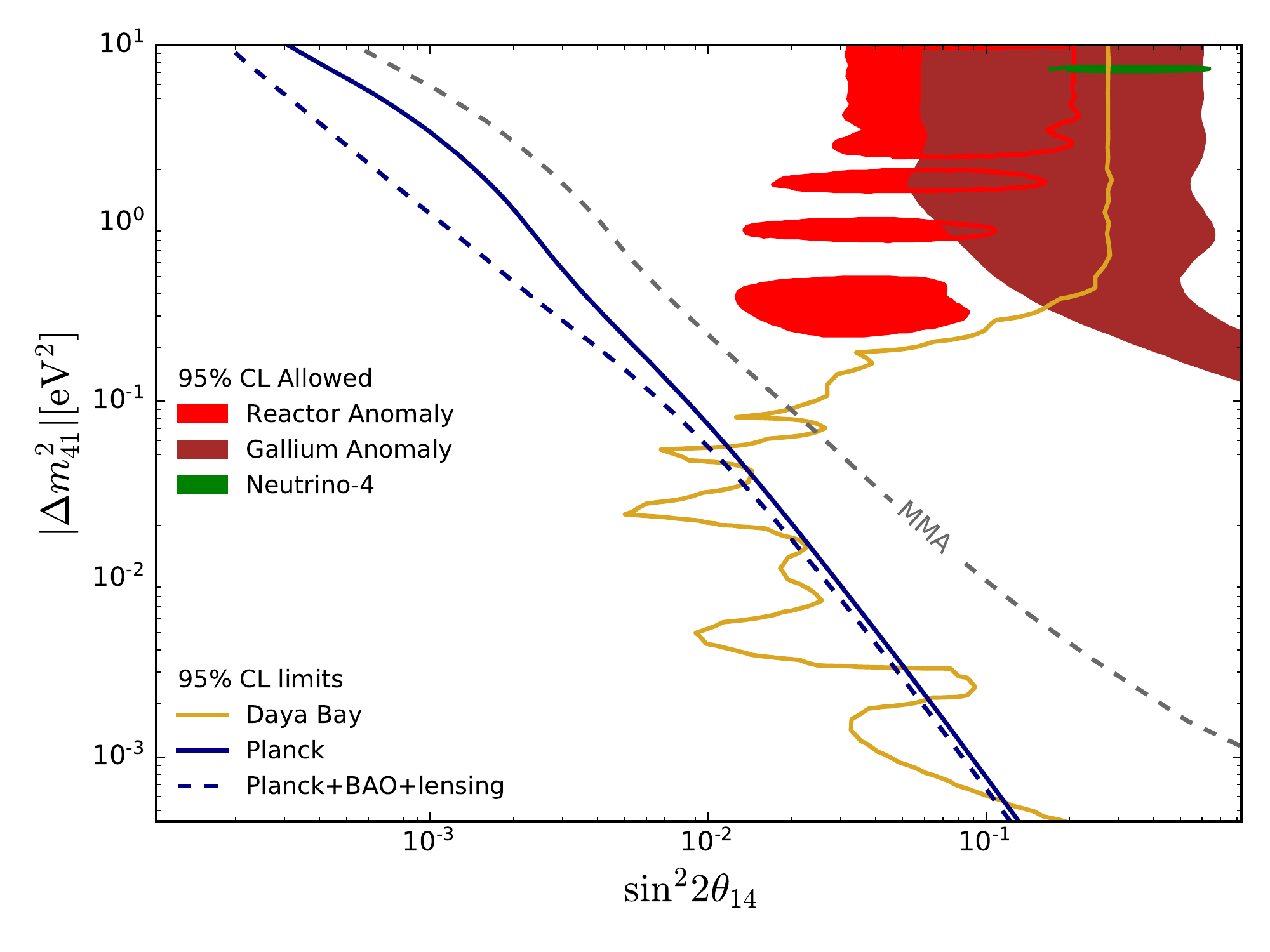}\label{fig:elec_dis_a}}
\subfigure[]{\includegraphics[width=0.99\columnwidth]{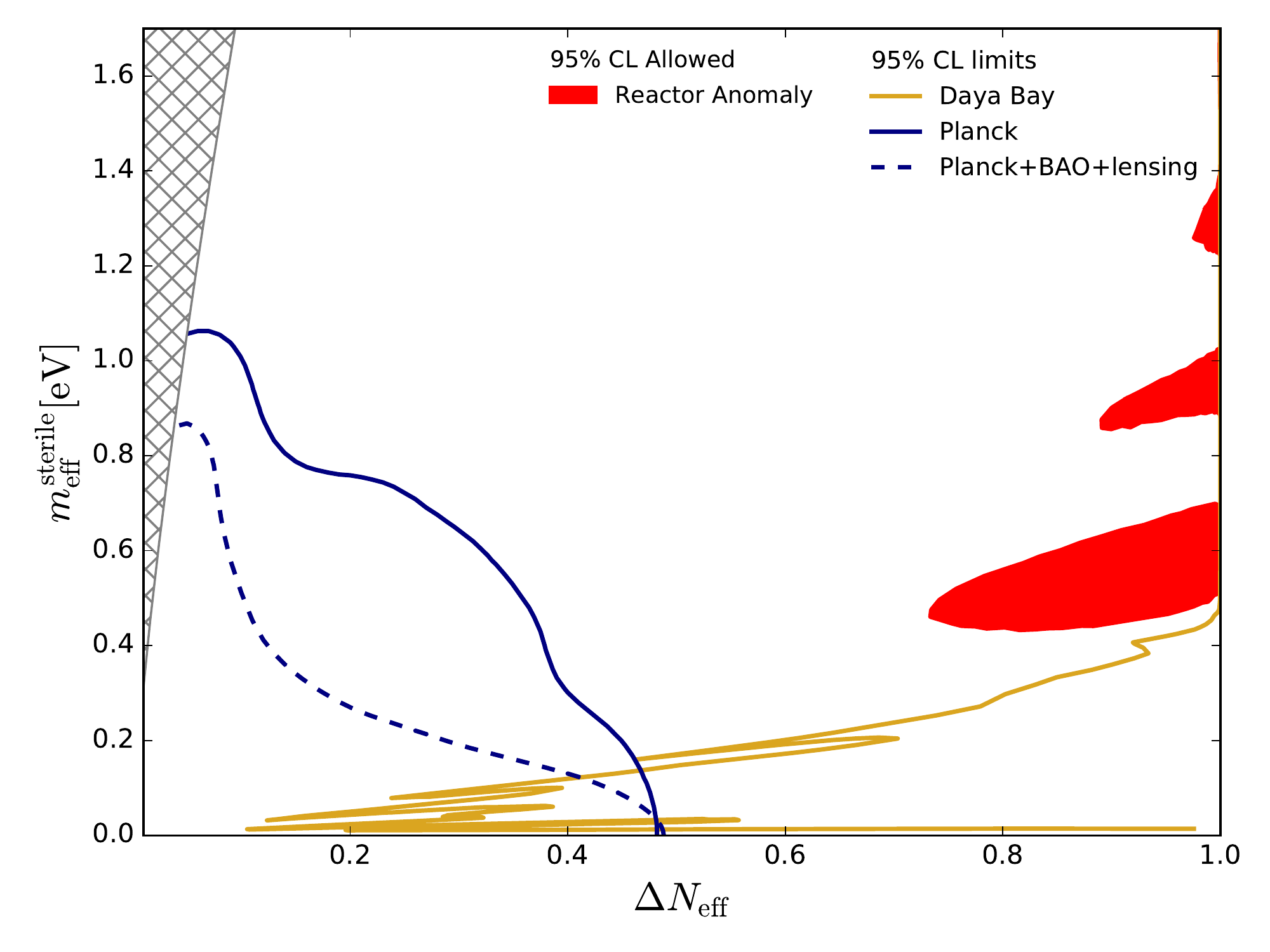}\label{fig:elec_dis_b}}
\caption{(a) shows, in the neutrino-oscillation parameter space, limits on the electron flavour mixing with the fourth mass state, using a $1+1$ model. The exclusion region from the Daya Bay oscillation experiment, and the allowed regions from the Neutrino-4 experiment and the reactor anomaly, come from searches for \nuebar disappearance. The allowed region from the gallium anomaly comes from a search for \nue disappearance. For the Daya Bay line, everything to the right is ruled out at 95\% CL. The solid blue line labeled `Planck' shows the exclusion using the Planck TT, TE, EE and low-multipole EE power spectra, using Eq.~\ref{eq:EffectiveMass} to relate $m_{\mathrm{eff}}^{\mathrm{sterile}}$ to $m_{4}$, with the region to the right ruled out at 95\% CL.
The dashed blue line shows the impact of further including CMB and BAO data into the Planck limit, again using Eq.~\ref{eq:EffectiveMass}. The dashed grey line illustrates the impact on the Planck limit (the solid blue) of using the mean momentum approximation (MMA). Graph (b) shows the same set of limits (minus the MMA line) in the cosmological parameter space. The Neutrino-4 and gallium-anomaly lines are no longer visible as they are compressed up along the $\Delta N_\mathrm{eff}=1$ axis. The hatched region corresponds to the prior of $m_{4}<\unit[10]{eV}$ assumed in the Planck analysis.}
\label{fig:electron_disappearance}
\end{figure*}

Limits from neutrino disappearance experiments can be interpreted in the $1+1$ model, which assumes that only one active flavour state mixes into the fourth mass state and that the three other mass states form a single, mass-degenerate state, $\nu_{d}$. For \nue disappearance experiments, we allow only the \nue flavour to mix into the fourth mass state. This is equivalent to varying $\theta_{14}$ whilst fixing $\theta_{24}=\theta_{34}=0$. In this model, we can write
\begin{eqnarray}
\nu_{e} &=& \cos\theta_{14}\nu_{d}-\sin\theta_{14}\nu_{4},\\
\nu_{s} &=& \sin\theta_{14}\nu_{d}+\cos\theta_{14}\nu_{4}.
\end{eqnarray}

LASAGNA calculates the Bloch vectors 
\begin{equation}
(P_{0},{\bf P})=(P_{0}, P_{x}, P_{y}, P_{z})
\end{equation}
 for neutrinos and  $(\overline{P}_{0}, \overline{\bf P})$ for anti-neutrinos using the $1+1$ model. The resulting vector $P_{s}^{+} = (P_{0}+\overline{P}_{0}) + (P_{z}+\overline{P}_{z})$ enters the expression
\begin{equation}
\Delta N_\mathrm{eff}=\frac{\int (p/T)^{3}(1+e^{p/T})^{-1}P_{s}^{+}\,\mathrm{d}(\frac{p}{T})}{4\int (p/T)^{3}(1+e^{p/T})^{-1}\,\mathrm{d}(\frac{p}{T})},
\end{equation}
where the momentum distribution, $p$, of the neutrinos is assumed to obey a Fermi-Dirac distribution at temperature $T$. A temperature range of $T=[40,1]\,\mathrm{MeV}$ covers the period from the beginning to the end of decoupling. We assume the lepton asymmetry, \mbox{$L=(n_{l}-n_{\overline{l}})/n_{\gamma}$}, to be zero. It was shown in Ref.~\cite{ref:Us} that the 
Planck exclusion region is significantly reduced
in a 1+1 model for $\nu_{\mu}$ disappearance for large lepton asymmetries ($L = 10^{-2}$).

We use LASAGNA to calculate $\Delta N_{\mathrm{eff}}$ for a grid in the oscillation parameter space of $|\Delta m^{2}_{41}| \equiv|m_{4}^{2}-m_{d}^{2}|$ and $\sin^{2}(2\theta_{14})$, as shown in Fig.~\ref{fig:translate_elec_dis_a}.
Equation~\ref{eq:EffectiveMass} allows us to express this result for all relevant combinations of $\Delta N_{\mathrm{eff}}$, $m_{\mathrm{eff}}^{\mathrm{sterile}}$, $\sin^{2}(2\theta_{14}),$ and $|\Delta m^{2}_{41}|$ (Figs.~\ref{fig:translate_elec_dis_b}--\ref{fig:translate_elec_dis_d}). 
\newtext{The figures show that the impact of the sterile state on $\Delta N_\mathrm{eff}$ is minimal for small $\sin^{2}(2\theta_{14})$ and $|\Delta m^{2}_{41}|$, increasing to a full extra degree of freedom, $\Delta N_\mathrm{eff}=1$, at larger values of the mixing angle and mass splitting. This is related to the amount of thermalisation of the fourth neutrino state in the early universe: a larger mixing angle allows a higher thermalisation rate, and a larger effective sterile neutrino mass (corresponding to a larger mass splitting) increases the temperature at which the thermalisation occurs. More explanation of this can be found in Refs.~\cite{ref:Hannestad2012,ref:MarkThomson}.}

In Fig.~\ref{fig:elec_dis_a}
we express the Planck exclusion limit in the parameter space $(\sin^{2}(2\theta_{14}), |\Delta m^{2}_{41}|)$ and overlay the limit from Daya Bay, and the allowed regions from Neutrino-4 and the gallium and reactor anomalies. 
The equivalent contours translated into the cosmological parameter space $(m_{\mathrm{eff}}^{\mathrm{sterile}}, \Delta N_{\mathrm{eff}})$ are shown in Fig.~\ref{fig:elec_dis_b}. In both figures, we show the Planck limit with and without the BAO and CMB lensing data. 

The limits obtained using the Planck data with and without the
BAO and CMB lensing data are strongly constraining in both parameter spaces in the region above $|\Delta m^{2}_{41}|^{2}\approx\unit[0.1]{eV^{2}}$ and $m_{\mathrm{eff}}^{\mathrm{sterile}}\approx\unit[0.2]{eV}$, and exclude the allowed regions from the Neutrino-4 experiment, and from the gallium and reactor anomalies. The Daya Bay experiment is sensitive to the regions of low $|\Delta m^{2}_{41}|$ and $m_{\mathrm{eff}}^{\mathrm{sterile}}$, where the cosmological data are less constraining.

\begin{figure*}[htbp]
\begin{center}
\subfigure[]{\includegraphics[width=1.02\columnwidth]{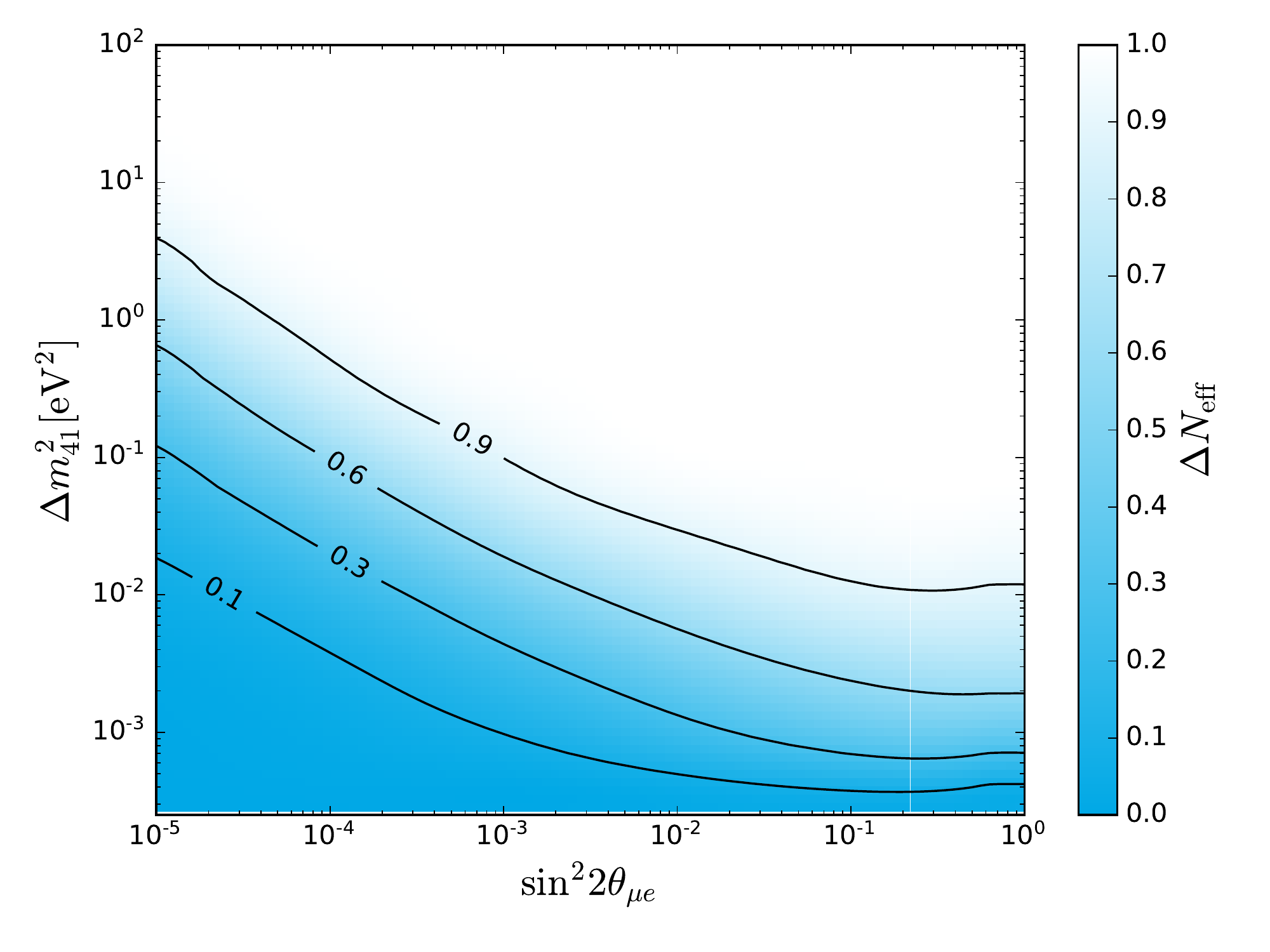}\label{fig:translate_elec_app_a}}
\subfigure[]{\includegraphics[width=1.02\columnwidth]{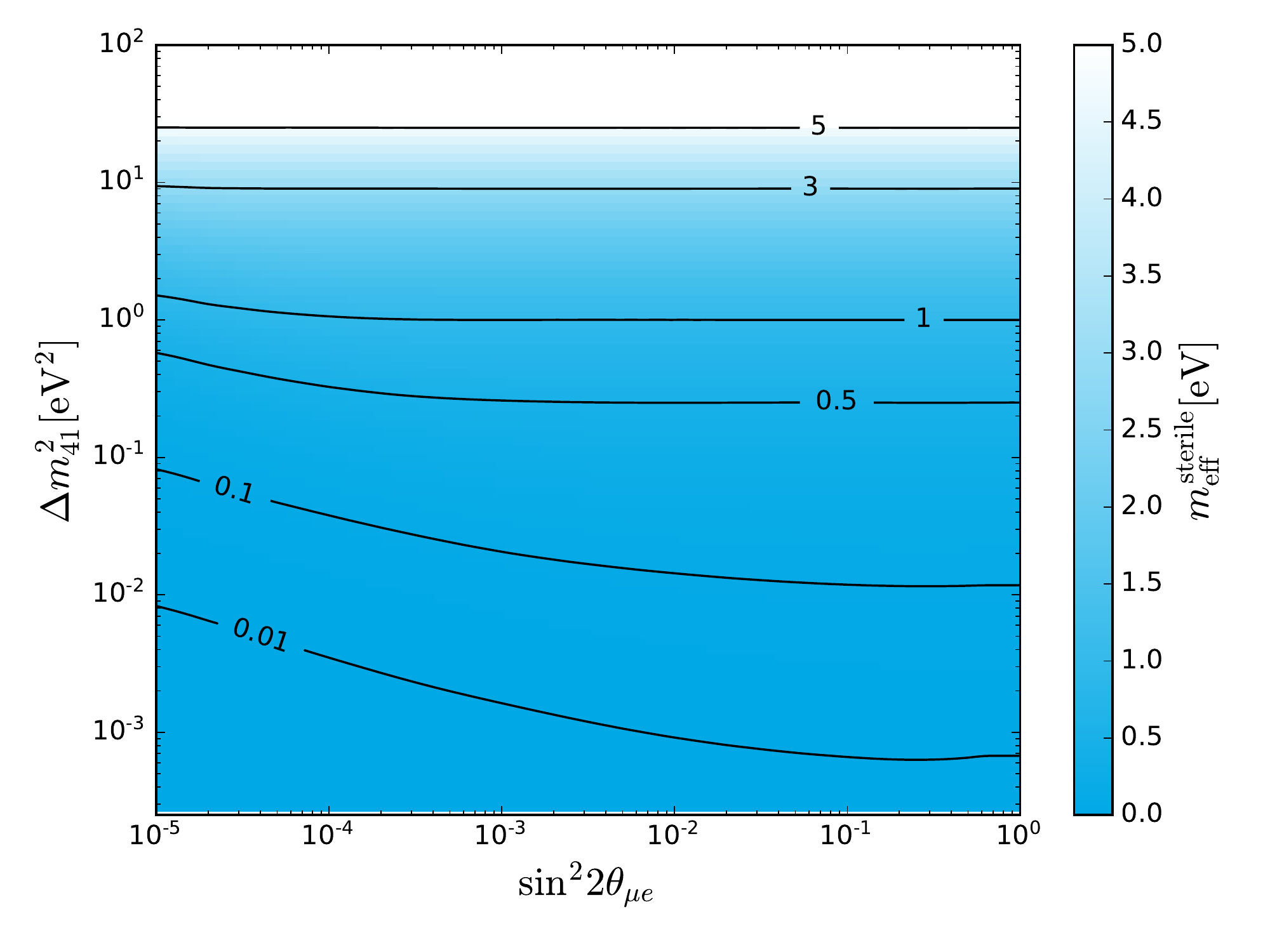}\label{fig:elec_app_b}}
\caption{The cosmological parameters (a) $\Delta N_{\rm eff}$ and (b) $m_\mathrm{eff}^\mathrm{sterile}$ calculated in the oscillation parameter space ($\Delta m^{2}_{41}$, $\sin^{2} (2 \theta_{\mu e}))$ using the mean momentum approximation as described in the text.}
\label{fig:translations_appearance}
\end{center}
\end{figure*}

\section{Electron neutrino appearance in a $3+1$ model}
\label{sec:nueappearance}

When considering $\sin^{2}(2\theta_{\mu e})$, both mixing angles $\theta_{14}$ and $\theta_{24}$ must be allowed to be non-zero to allow both \nue and \numu flavours to mix into the $\nu_{4}$ state, and so we work in the $3+1$ model with one sterile and three active neutrino flavours, albeit setting $\theta_{34}=0$. 
This model can be solved exactly~\cite{ref:Gariazzo} but working with the full momentum dependence of the quantum kinetic equations is computationally very intensive. Hence, we use the mean momentum approximation (MMA) following the prescription of Ref.~\cite{ref:mma} summarized below.

The neutrino density matrix,
\begin{equation}
    \rho(x,y)=\left(
    \begin{array}{cccc}
    \rho_{ee} & \rho_{e\mu} & \rho_{e\tau} & \rho_{es}\\
    \rho_{\mu e} & \rho_{\mu\mu} & \rho_{\mu\tau} & \rho_{\mu s}\\
    \rho_{\tau e} & \rho_{\tau\mu} & \rho_{\tau\tau} & \rho_{\tau s}\\
    \rho_{se} & \rho_{s\mu} & \rho_{s\tau} & \rho_{ss}\\
    \end{array}
    \right),
\end{equation}
depends on the mixing angles and mass splittings. It can be written as a function of reduced time, $x\equiv m/T$, and reduced momentum, $y\equiv p/T$, where $m$ is an arbitrary mass scale and $T$ is the initial temperature of the thermal, active neutrinos. This matrix is used to calculate $\Delta N_{\mathrm{eff}}$ for any required values of $\theta_{14}$, $\theta_{24}$ and $\Delta m^{2}_{41}$ as
\begin{equation}
    \Delta N_{\mathrm{eff}}=\frac{1}{2}\left(\mathrm{Tr}(\rho)+\mathrm{Tr}(\overline{\rho})-6\right).
\end{equation}
The MMA assumes that the momentum dependence of $\rho(x,y)$ can be factorized out as a Fermi-Dirac distribution, $\rho(x,y)\rightarrow f_{\mathrm{FD}}(y)\rho(x)$. The equations of motion for the neutrino and anti-neutrino density matrices is then written assuming that all neutrinos have the same momentum, $\left<y\right>$.

\begin{figure}[htbp]
    \centering
    \includegraphics[width=0.5\textwidth]{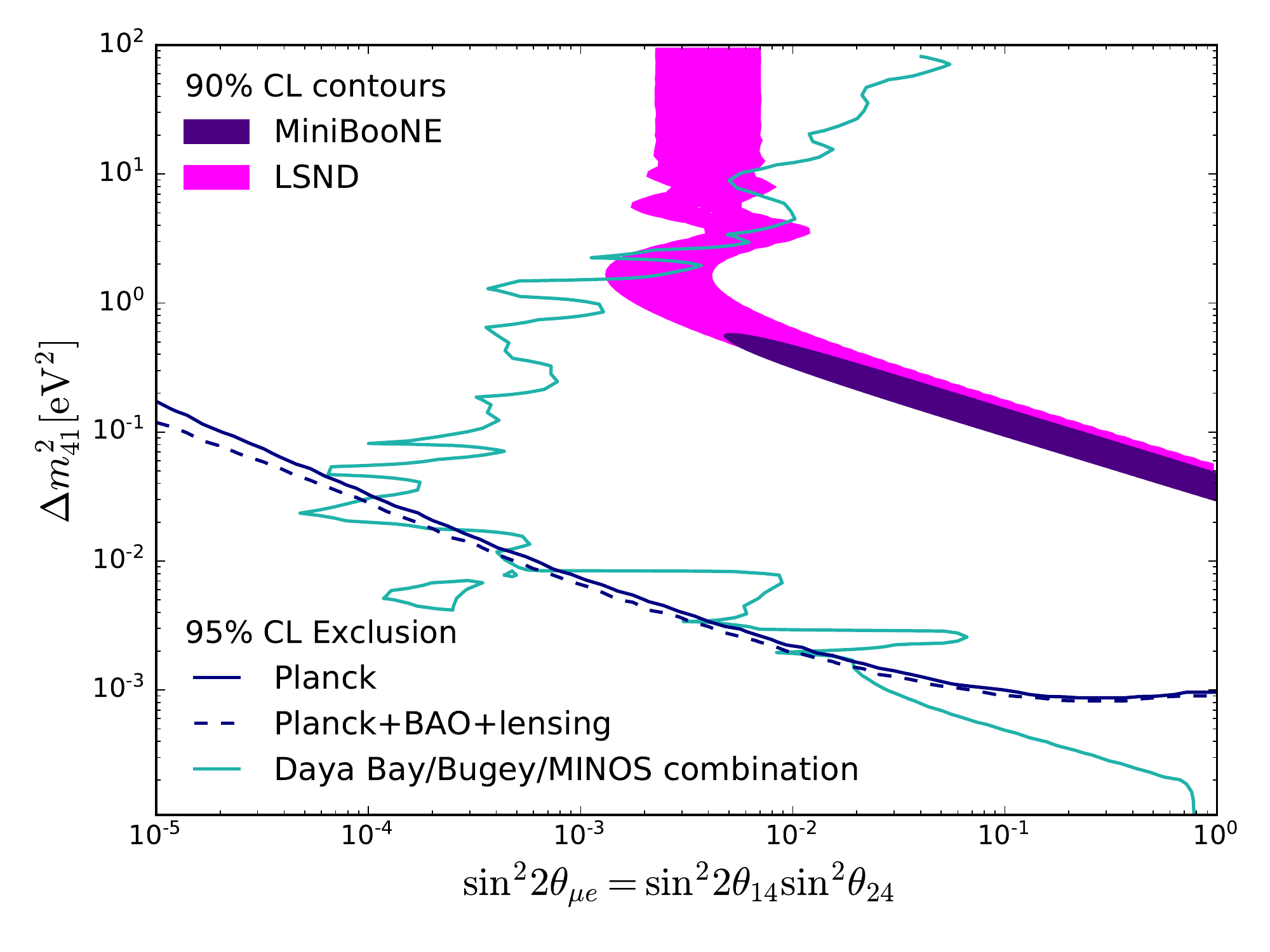}
    \caption{Limits on the parameters governing $\nu_\mu\rightarrow\nu_e$ appearance in a 3+1 model, shown in the neutrino-oscillation parameter space. Solid regions are the allowed regions from the MiniBooNE and LSND measurements. The light blue line is an exclusion region from the Daya Bay/Bugey/MINOS combined analysis. The dark blue lines show the Planck exclusion region, expressed in this parameter space, with (dashed) and without (solid) the BAO and CMB lensing data included.}
    \label{fig:appearance}
\end{figure}

We solve the resulting differential equations of motion numerically with an implicit Runge-Kutta algorithm of order $5$, RADAU5~\cite{ref:RADAU5}, using a publicly available C++ implementation~\cite{ref:RungeKuttaImplementation}.
To evaluate $\Delta N_{\mathrm{eff}}$, we evolve the density matrix from $T=\unit[100]{MeV}$ to $T=\unit[1]{MeV}$. To project the cosmological limits onto the $\sin^{2}(2\theta_{\mu e})$ axis, we minimise the value of $\Delta N_{\mathrm{eff}}$ as a function of $\theta_{14}$ and $\theta_{24}$ along a contour of constant $\sin^{2}(2\theta_{\mu e})$; the derived 95\% confidence limits therefore assume the maximum possible thermalisation for a given value of $\theta_{\mu e}$. The resulting values of $\Delta N_\mathrm{eff}$ as a function of $\Delta m^{2}_{41}$ and $\sin^{2}(2\theta_{\mu e})$ are shown in Fig~\ref{fig:translations_appearance}.

In the region $|\Delta m^{2}_{41}|\lesssim|\Delta m^{2}_{31}|$, the mass splitting $\Delta m^{2}_{41}$ is driving neutrino oscillations at wavelengths similar to those driven by the active-neutrino mass splittings. This is referred to as the degenerate region, and in this region the RADAU5 solver slows down drastically due to the stiffness of the problem when degeneracies are crossed. To mitigate this, we increase the tolerance by a factor of 10 after every 100,000 steps of the algorithm, starting from a default tolerance of $10^{-10}$, reaching a maximum tolerance of $10^{-4}$ required for certain parameters to converge quickly.

We evaluate the impact of the MMA by repeating the \nue disappearance analysis in the $1+1$ model using this approximation. The result of this is shown in Fig~\ref{fig:elec_dis_a}, illustrating that, under the MMA, the cosmological exclusion contours expressed in the $(\Delta m^{2}_{41},\sin^{2}(2\theta_{14}))$ parameter space become slightly weaker.

In Figure~\ref{fig:appearance}, we show the Planck exclusion contours, with and without the BAO and CMB lensing data, in the $(\Delta m^{2}_{41},\sin^{2}(2\theta_{\mu e}))$ parameter space. We compare this to the limits from the Daya Bay/Bugey/MINOS combination, and the allowed regions from the LSND and MiniBooNE $\nue\rightarrow\numu$ searches. The Planck exclusion region strongly excludes the entirety of the LSND and MiniBooNE allowed regions. The Daya Bay/Bugey/MINOS combined exclusion region is comparable in its exclusion power to that from the Planck data for mass splittings below $\Delta m^{2}_{41}\approx\unit[5\times 10^{-2}]{eV^{2}}$ and becomes more constraining below $\Delta m^{2}_{41}\approx\unit[10^{-3}]{eV^{2}}$.
\section{Conclusions}
\label{sec:conclusions}

The discovery of a sterile neutrino would have major implications for the field of particle physics. The presence of both possible observations from neutrino oscillation experiments such as LSND and MiniBooNE, negative results from other oscillation experiments, and negative results from cosmological experiments, have left the field in an ambiguous situation. A particular challenge in drawing conclusions is quantitative comparison of limits from neutrino oscillation data with those from cosmology, due to the different parameter spaces in which measurements from these two sets are expressed.

In this article, we discuss a procedure to convert limits on sterile neutrinos between the $(|\Delta m^{2}_{41}|,\theta_{14},\theta_{24})$ parameter space of neutrino oscillation physics and the $(m^{\mathrm{sterile}}_{\mathrm{eff}},\Delta N_{\mathrm{eff}})$ parameter space of cosmology. We use the LASAGNA software package to solve the quantum kinetic equations of neutrinos in the early universe in a $1+1$ model, allowing us to compare the exclusion regions obtained from Planck data with both allowed regions and exclusion regions from \nue and \nuebar disappearance searches. In a $3+1$ model, we use a mean momentum approximation to solve the quantum kinetic equations, allowing us to compare the Planck exclusion with allowed regions and exclusion regions corresponding to $\numu\rightarrow\nue$ searches. We find that the Planck data strongly excludes the allowed regions from the Neutrino-4, LSND and MiniBooNE experiments, as well as from the gallium and reactor anomalies. Compared to the Daya Bay exclusion region from \nue disappearance, Planck is much more constraining above $|\Delta m^{2}_{41}|\approx\unit[0.1]{eV^{2}}$ and $m^\mathrm{sterile}_\mathrm{eff}\approx\unit[0.2]{eV}$, whereas at lower values, Daya Bay provides a more stringent exclusion on $\theta_{14}$. The Planck data provide the strongest exclusion on the $\theta_{\mu e}$ parameter that describes $\numu\rightarrow\nue$ appearance above $\Delta m^{2}_{41}\approx\unit[5\times 10^{-2}]{eV^{2}}$; below this value, the Daya Bay/Bugey/MINOS combination becomes comparable in terms of its exclusion power.

\newtext{Experimental and theoretical efforts are ongoing to relieve the tension between positive signals from appearance experiments and the strong exclusions from disappearance measurements and cosmology. Appearance experiments such as MicroBooNE~\cite{ref:MicroBooNE} and the SBN programme~\cite{ref:SBNProgramme} have the potential to rule out or confirm the previous appearance signals. Theoretical work on the cosmological side has to limit thermalisation of the sterile neutrino state in order to maintain $N_\mathrm{eff}\approx 3$. Examples include the introduction of new interactions for the sterile neutrino~\cite{ref:HannestadNonStandardInteractions,ref:ArchidiaconoNonStandardInteractions,ref:SavianoNonStandardInteractions,ref:ChuNonStandardInteractions}, a large lepton-antilepton asymmetry in the early universe~\cite{ref:ChuLeptonAsymmetry,ref:FootLeptonAsymmetry,ref:SavianoLeptonAsymmetry}, and the introduction of reheating at low temperatures~\cite{ref:SalasLowReheating,ref:KawasakiLowReheating,ref:GelminiLowReheating}.}

\begin{acknowledgements}
We are grateful
to Thomas Tram (ICG Ports\-mouth) for help in running the LASAGNA code. We thank Joe Zuntz and Richard Battye (Manchester), and Steen Hannestad (Aarhus) for helpful discussions. This work has been supported by the Science and Technology Facilities Council, part of UK Research and Innovation, the Royal Society, and the European Research Council. Participation of one of the authors (P.G.) has been funded from the European Union's Horizon 2020 research and innovation programme under the Marie Sk\l{}odowska-Curie grant agreement no.\ 752309.
\end{acknowledgements}

\bibliographystyle{epjc}
\bibliography{main.bib}

\end{document}